\documentclass{mn2e}

\usepackage{psfig,subfigure,graphicx,mncite}

\newcommand{\bvec}[1]{\mbox{\boldmath ${#1}$}}
\newcommand{\deriv}[2]{\mbox{${{\displaystyle d#1}\over {\displaystyle d#2}}$}}
\newcommand{\pderiv}[2]{\mbox{${{\displaystyle\partial#1}\over {\displaystyle\partial#2}}$}}

\begin{document}

\title[Slingshot prominences]
{Slingshot prominences above stellar X-ray coronae}

\author
[M. Jardine, \& A.A. van Ballegooijen]
{M. Jardine$^1$, 
\thanks{E-mail: moira.jardine@st-and.ac.uk}
\& A.A. van Ballegooijen$^2$ 
\\
$^1$School of Physics and Astronomy, Univ.\ of St~Andrews, 
St~Andrews, 
Scotland KY16 9SS \\
$^2$Harvard-Smithsonian Centre for Astrophysics, 60 Garden Street, Cambridge, M.A. 02138 USA \\
\\
} 

\date{Received; accepted 2005}

\maketitle

\begin{abstract}

We present a new model for the coronal structure of rapidly rotating
solar-type stars.  The presence of prominences trapped in co-rotation
2 to 5 stellar radii above the stellar surface has been taken as evidence that the
coronae of these stars must be very extended.  The observed surface
magnetic fields, however, cannot contain X-ray emitting gas out to
these distances.  We present an alternative model: that these
prominences are trapped in long thin loops embedded not in the X-ray
corona, but in the wind.  Above coronal helmet streamers,
oppositely-directed wind-bearing field lines reconnect to form closed loops which
then fill up with gas that was originally part of the wind.  We demonstrate
that static equilibria exist for these loops at a range of pressures
and temperatures.  The maximum loop height falls as the rotation rate increases, but rises as the loop temperature decreases.  For a solar-mass star with rotation period 0.5 days, whose X-ray corona
extends 1R$_{\star}$ above the surface, loops at temperatures of $10^{4}$K can
extend out to 5R$_{\star}$.

\end{abstract}


\section{Introduction}

Since the days of the first X-ray satellites, the X-ray emission from solar-like stars has been a puzzle. For main-sequence stars it is a clear function of rotation rate, rising steeply for stars rotating faster than the Sun until  rotation rates of about $15-20$kms$^{-1}$ where it reaches a plateau at $L_{x}/L_{bol}=10^{-3}$ \cite{vilhu84}.  Beyond this rotation rate is the ``saturated'' regime where the X-ray luminosity is independent of rotation rate. This ``saturated'' behaviour persists until rotation rates of about $vsini>100$kms$^{-1 }$, where the X-ray luminosity begins to decrease
again.  This regime is referred to as ``supersaturated'' \cite{prosser96,randich98}. 

The initial increase seems linked to the enhanced dynamo activity that accompanies increased rotation, but the causes of the so-called saturation and supersaturation are still a matter of debate. While dynamo activity within the convective region may saturate when the back-reaction of the fluid becomes significant,  the X-ray emission produced when this field escapes into the corona is only an indirect measure of this process. Levels of X-ray emission may be determined by the physics of the corona as much as by the physics of sub-surface dynamo activity. Saturation of X-ray emission may represent the inability of the surface, or the corona, to accommodate any more flux  \cite{vilhu84}, an inability of the energy release process to extract more energy from the magnetic field, or centrifugal stripping of the corona at high rotation rates \cite{jardine99stripping}.

A further puzzle is the nature of this enhanced emission from the rapid rotators. While early observations showing little rotational modulation of the X-ray emission indicated that the emission (and hence the coronae) must be very extended, some observations suggested that the emission came from compact (solar-like) regions at high latitudes on the stars where they did not suffer from rotational self-eclipse \cite{singh96,siarkowski96,giampapa96,jeffries98}. In this case, however, a high density was required to explain the high X-ray luminosity. In the last few years, however, spectra obtained with FUSE, Chandra and XMM-Newton have made a significant impact on this problem by revealing high coronal densities \cite{dupree93,schrijver95,brickhouse98,audard01,mewe01,young01,gudel01XMM,sanz_forcada_abdor_03,sanz_forcada_03}. A critical BeppoSAX observation of two flares on the rapid rotator AB Dor showed no rotational modulation of the flare decay phases, which lasted for more than one stellar rotation \cite{maggio2000}. Flare modelling suggested the flaring regions were compact (0.3$R_\star$)  and therefore must have been located at high latitudes. The ability to locate emitting features is improving rapidly with the beginnings of ``X-ray Doppler imaging" and further eclipse observations \cite{brickhouse200144iboo,gudel03aCrB}. Most recently, simultaneous X-ray spectra and surface magnetograms (obtained through Zeeman-Doppler imaging) have made it possible to couple the surface field to the coronal emission using field extrapolation techniques \cite{jardine02structure,hussain2001potential,hussain_current_02,jardine02xray}.
A consistent picture appears to be emerging of the coronae of these rapid rotators. They are compact and highly structured, with discrete emitting regions often situated at the high latitudes where Doppler imaging shows many of the large star spots to be located. The coronal densities are typically higher than on the Sun, giving the very high X-ray luminosities. 

One type of observation, however, appears to be quite inconsistent with this picture. On almost all these rapid rotators with $H_\alpha$ observations, slingshot prominences are detected  \cite{cameron89cloud,cameron89eject,cameron92alpper,jeffries93,byrne96hkaqr,eibe98re1816,barnes20PZTel,donati20RXJ}. These are observed as transient absorption features that move through the H$_\alpha$ line profiles. In many instances these features re-appear on subsequent stellar rotations, often with some change in the time taken to travel through the line profile. These features are interpreted as arising from the presence of clouds of cool, dense gas co-rotating with the star and confined within its outer atmosphere. As many as six may be present in the observable hemisphere.  What is most surprising about them is their location, which is inferred from the  time taken for the absorption features to travel through the line profile. Values of several stellar radii from the stellar rotation axis are typically found, suggesting that the confinement of these clouds is enforced out to very large distances. Indeed the preferred location of these prominences appears to be at or beyond the equatorial stellar co-rotation radius, where the inward pull of gravity is exactly balanced by the outward pull of centrifugal forces. Beyond this point, the effective gravity (including the centrifugal acceleration) points outwards and the presence of a restraining force, such as the tension in a closed magnetic loop, is required to hold the prominence in place against centrifugal ejection. The presence of these prominences therefore immediately requires that the star have many closed loop systems that extend out for many stellar radii.

While these observations may lend support to the hypothesis that these stars have coronae that are very extended, it does not imply that they are smooth. Of all the multipolar components of a coronal field, the dipole component falls of most slowly with height, and so it is natural to suppose that a very extended corona will resemble a dipole field. A smooth and symmetric field like this, however, would give prominence locations which were, by symmetry, all in the equatorial plane and uniformly spread around the star \cite{jardine2001eqm}. Prominences in the equatorial plane of a star are unlikely to be observed, however,  unless the inclination of the stellar rotation axis is close to $90^\circ$. For a star such as AB Dor, whose rotation axis is inclined at some $60^\circ$ to the observer, prominences that are located at around $3 R_\star$ from the rotation axis must be located at high latitudes in order to transit the disc. The large number of prominences observed (typically 6 in the observable hemisphere for AB Dor) suggest that there is significant azimuthal structure in the field supporting the prominences even at several stellar radii from the rotation axis. 

\begin{figure}
\begin{center}

\includegraphics[width=3in]{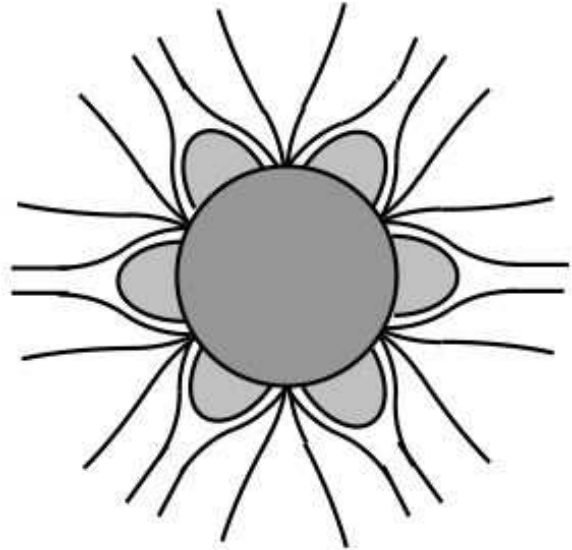}

 \caption{A schematic diagram of the magnetic field in the equatorial plane of the star as seen by an observer looking along the rotation axis. Wind-bearing, open field lines emerge from between regions of closed field that are bright in X-rays. Above these closed field regions, helmet streamers form. It is above the cusps of these helmet  streamers that prominence-bearing loops may form.}

  \label{polarview}
 \end{center}
\end{figure}
\begin{figure*}
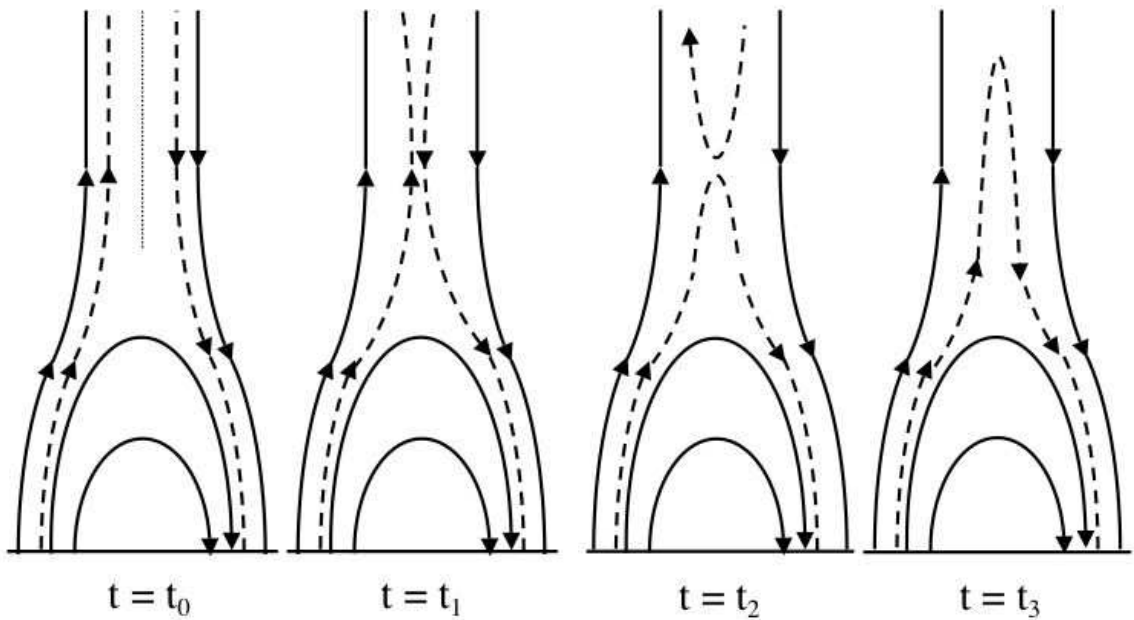

\begin{center}

\includegraphics[width=3in]{jardinefig2a.epsf}
\includegraphics[width=3in]{jardinefig2b.epsf}

 \caption{A schematic diagram of the formation of prominence-bearing loops. Initially, at $t=t_0$ a current sheet is present above the cusp of a helmet streamer. Reconnection at in the current sheet at $t=t_1$ produces a closed loop at $t=t_2$. The stellar wind continues to flow until pressure balance is restored, thus increasing the density in the top of this new loop. Increased radiative losses cause the loop to cool and the change in internal pressure forces it to a new equilibrium at $t=t_3$.}

  \label{cartoon}
 \end{center}
\end{figure*}

This high degree of complexity in the coronal field is supported by the degree of complexity seem in the surface magnetograms acquired by Zeeman-Doppler imaging and by traditional Doppler imaging \cite{donati97abdor95,donati99abdor96}. Multipolar flux is seen at all latitudes on AB Dor, even at very high latitudes. One region where it is not possible to determine the field polarity is, however, very close to the rotation pole where the Zeeman signature of often suppressed by the presence of the polar spot that is typically found on rapid rotators \cite{strassmeier96table}. By modelling the effect on the coronal structure of placing a unipolar field in this dark polar cap, \scite{jardine2001eqm,mcivor03polar} have shown that the resulting preferred prominence locations would all be in the equatorial plane. This suggests that the polar field must be of mixed polarity.

Even without the large-scale smoothing effect of a dipolar field component hidden in the dark polar caps, however, the global field topology implied by the surface magnetograms does not appear to be consistent with the large number of observed prominence locations. The coronal structure of both AB Dor and LQ Hya has been studied by extrapolating the coronal magnetic field from the surface magnetograms \cite{jardine02structure,mcivor04lqhya}. While the fields close to the surface may be highly structured, at heights of several stellar radii where the AB Dor prominence systems are found, the field is largely smooth and resembles a highly-inclined dipole. This would give only two longitude bands in which prominences would be found. By filling these magnetic loops with hydrostatic atmospheres, \cite{jardine02xray} have determined the coronal density and X-ray emission and shown them to be consistent in magnitude and rotational moduation with the observations. Comparison with simultaneous Chandra spectra also gives consistent results \cite{hussain_chandra1_05}. The X-ray emission typically comes from the high-latitude regions where the large flux concentrations are found and is confined within $1R_\star$ of the surface. Above these heights, the pressure of the hot coronal gas is sufficient to break open the magnetic flux loops and escape to form the stellar wind. 

This presents a serious dilemna for models of prominences which require the restraining force of closed magnetic field lines to contain them. Such large loops cannot contain the ten million degree plasma of the stellar corona at the observed prominence heights of several stellar radii. We can see the same effect at work on the Sun.  The distance at which the field becomes open can be estimated by comparing the gas pressure (assuming an isothermal, hydrostatic atmosphere) with the magnetic pressure of, say, a dipolar field. For typical solar parameters, this distance is about a solar radius above the surface. Above this height, the coronal gas is capable of pushing open the field lines and escaping as the solar wind. 

It seems then that the picture of compact, highly-structured, dense coronae that emerges from X-ray observations and from surface magnetograms coupled with field extrapolations is at odds with the presence of many individual prominences trapped in the coronae of these stars at distances of 2-5 $R_\star$ above the surface. The aim of this paper is to present an alternative model for stellar coronae that is consistent with all of these observations. We suggest that the prominences are embedded not in the closed, X-ray emitting corona, but in the stellar wind.  As on the Sun, large helmet streamers are likely to lie above active regions on the surface (see Fig. \ref{polarview}). Above the cusps of these helmet streamers, current sheets form, separating opposite field polarities. Such current sheets are unstable to tearing and subsequently reconnection, which will form closed loops. These loops will be filled up by the stellar wind which will continue to flow until a sufficient back-pressure develops in the loop. If the loop summit reaches above the co-rotation radius, then the outward pull of centrifugal forces may be enough to balance the inward pull of magnetic tension in the loop. If a mechanical equilibrium can be found, then the loop may be supported within the wind.

The aim of this paper is to demonstrate that such cool loops, with their summits above the co-rotation radius, can reach an equilibrium.

\section{Method}

We take a simple model shown schematically in the left panel of Fig. \ref{cartoon} where the background field is closed up to some height $y_s$ and then open beyond that height. This height is traditionally referred to as the {\em source surface} \cite{altschuler69}. We then consider the shape adopted by a closed loop that is embedded in this background field as shown in the rightmost panel of Fig. \ref{cartoon}. We consider a loop that is slender \cite{spruit81a} so that we can neglect any variation in pressure or density across its width. We also assume for simplicity that the loop is isothermal.
\begin{figure}
\begin{center}

\includegraphics[width=3in]{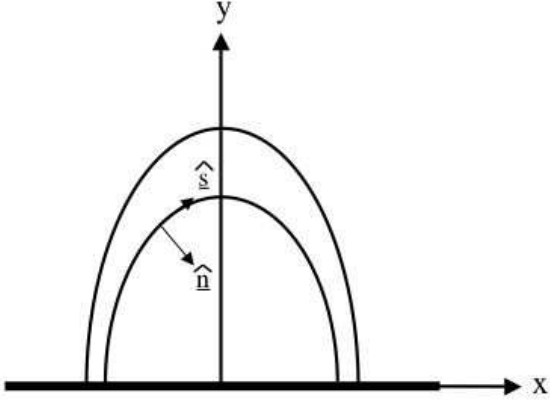}

 \caption{Schematic diagram of the coordinate system used.}

  \label{coords}
 \end{center}
\end{figure}

At any point along this loop, pressure balance at the loop boundary ensures that
\begin{equation}
    p_{i}+ \frac{B_i^2}{2\mu}=p_{e}+\frac{B_e^2}{2\mu}
\label{press_bal}
\end{equation}
where subscripts $i$ and $e$ refer to quantities internal and external to the loop respectively. 
The equilibrium shape of the loop is determined by the forces acting on it: pressure gradients, gravity and the Lorenz force:
\begin{equation}
    \bvec{\nabla p} -  \rho \bvec{g} = -\bvec{\nabla} \frac{B^{2}}{2\mu}                          
                     +(\bvec{B}\cdot\bvec{\nabla})\frac{\bvec{B}}{\mu}.
\label{eofm}
\end{equation}
We decompose this equation into components along (\bvec{\hat{s}}) and normal to (\bvec{\hat{n}}) the field (see Fig. \ref{coords}). We choose to place our loops in the equatorial plane of the star where the gravitational acceleration (including the effect of rotation) is purely radial. In the cartesian coordinate system shown in Fig. (\ref{coords}) we therefore have ${\bf g} = (0,g)$ and so we have
\begin{equation}
g = -\frac{GM_{\star}}{(y+R_\star)^{2}} + \omega^{2}(y+R_\star)    
\label{gravity}                 
\end{equation} 
where $y$ is the height above the stellar surface.
Along the direction of the loop, force balance is very simple since  the Lorentz force has no net component along the field. This determines the variation of the plasma pressure:
\begin{equation}
    \deriv{p}{s} = \rho g_{s}
\label{dpbyds}
\end{equation}
where $g_{s} = {\bf g.B}/|{\bf B}|$ is the component of
gravity (allowing for rotation) along the field. This reduces directly to
\begin{equation}
    \deriv{p}{y} = \rho g(y).
 \label{dpbydy}
\end{equation}
Hence along each field line the plasma simply adjusts itself into a 
hydrostatic balance with 
\begin{eqnarray}
    p & = & p_{0}\exp\left[\frac{m}{kT}\int g dy\right] \\
      & = & p_0 \exp\left[
                         \frac{m}{kT}\left(
                            \frac{-GM_{\star}}{R_\star}\frac{y}{y+R_\star}
                         + \frac{\omega^{2}}{2}y(y+2R_\star)
                                           \right) 
                           \right ]              .
      \end{eqnarray}
We note that the pressure $p_0$ at the base of each field line need not be the same, but for simplicity we choose one value $p_{0e}$ for the external field and one value $p_{0i}$ for the internal field.
\begin{figure}
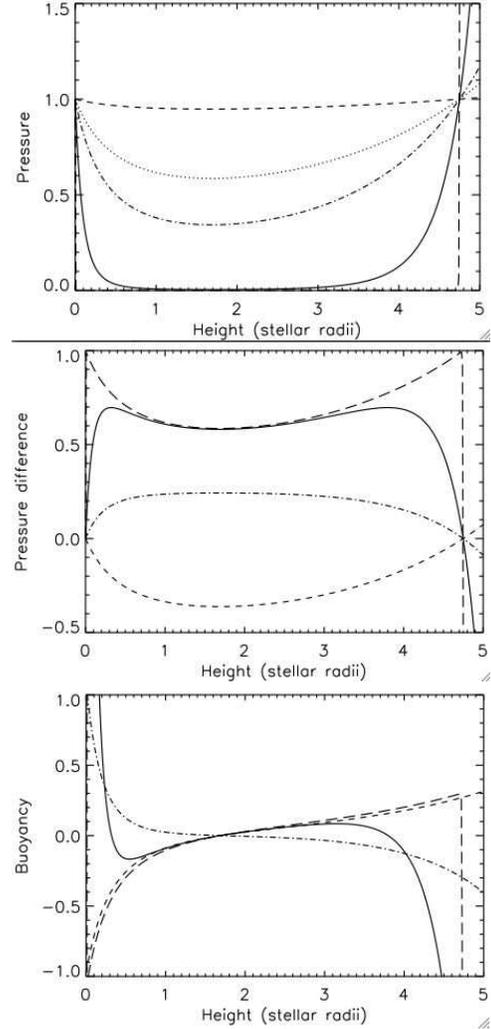

\begin{center}

\includegraphics[width=2.5in]{jardinefig4a.epsf}
\includegraphics[width=2.5in]{jardinefig4b.epsf}
\includegraphics[width=2.5in]{jardinefig4c.epsf}

 \caption{Variation with height of the plasma pressure scaled to its base value (top); the difference between the external and internal plasma pressures (middle) and the gradient of the pressure difference $\partial{(p_e-p_i)}/\partial{y}=(\rho_e-\rho_i)g$ i.e. the buoyancy (bottom) for loops that are cooler or hotter than their environment: $T_i=10T_e$ (dashes), $T_i=T_e$ (dots), $T_i=0.5T_e$ (dot-dashes), $T_i=0.1T_e$ (solid), $T_i=10^{-3}T_e$ (long dashes). For the very coolest loop (long dashes) the rapid drop in pressure near the surface and its steep rise at around 4.8$R_\star$ can be seen as near-vertical lines in the top panel. The temperature of the external medium is $T_e=2\times 10^7$K. Note that the pressure reaches a minimum at the co-rotation radius, which here is at $1.7 R_\star$ above the stellar surface.}

  \label{pressure}
 \end{center}
\end{figure}
\begin{figure}
\begin{center}

\includegraphics[width=2.7in]{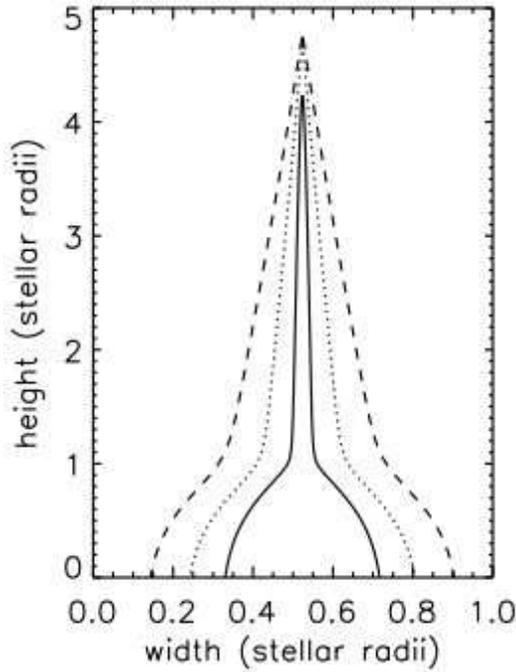}

 \caption{The shape taken up by cool field lines at temperatures of $5\times 10^6$K (solid), $2\times 10^5$K (dashed) and $2\times 10^4$K (dotted). Within the external field, the temperature is $T_e=2\times 10^7$K, the source surface $y_s=1$ and we choose $k=3$ so that each helmet streamer of the external field spans 60$^\circ$ of longitude at the stellar surface. The pressure at the loop base is twice that of the external field.}

  \label{single_shape}
 \end{center}
\end{figure}
For simplicity we take the external pressure to be in hydrostatic equilibrium and neglect the effect of the stellar wind. The balance of forces normal to the loop then determines the equilibrium shape of the loop. We define unit vectors along and normal to the field as:
\begin{eqnarray}
\bvec{\hat{s}} &=& \frac{(1,y')} 
{[1+(y')^{2}]^{1/2}} \\
    \bvec{\hat{n}} &=& \frac{(y',-1)}
{[1+(y')^{2}]^{1/2}},
\end{eqnarray}
where $y(x)$ defines the path of a flux tube and primes denote derivatives with respect to $x$. Gradients normal to the field can then be expressed as
\begin{equation}
 \bvec{\hat{n}} \cdot \bvec{\nabla} = \frac{1}{[1+(y')^{2}]^{1/2}} 
                               \left(y'\pderiv{}{x} -\pderiv{}{y} \right).
\end{equation}
If we also note that in (\ref{eofm}) the component of the magnetic tension term normal to the flux tube can be written as
\begin{equation}
\left[ (\bvec{B_i}\cdot\bvec{\nabla})\frac{\bvec{B_i}}{\mu} \right]_{\hat{n}}
       = \frac{-y''}{[1+(y')^{2}]^{3/2}}\frac{B^2_i}{\mu} \bvec{\hat{n}}
\end{equation}
then the component of force balance normal to the loop gives
\begin{equation}
-\pderiv{p}{y} = -\left[
                          \left(
                                  y'\pderiv{}{x} - \pderiv{}{y}
                            \right) \frac{B_i^2}{2\mu}
                         + \frac{y''}{1+(y')^2} \frac{B_i^2}{\mu}
                       \right]
                      - \rho g.
\end{equation}
In taking derivatives of $B_i^2$ it should be remembered that $B_i^2  = B_e^2 + 2 \mu (p_{e}- p_i)$.
Using (\ref{dpbydy}), however, this can be reduced to
\begin{equation}
\left(
       y'\pderiv{}{x} - \pderiv{}{y}
\right) \frac{B_i^2}{2}
         = -\frac{y''}{1+(y')^2} B_i^2.
         \label{shape}
\end{equation}
We choose to define as boundary conditions that the height of the loop (H) is given and also that the loop is flat at the top, i.e. $y(0) = H$ and $y'(0)=0$.

\section{Nature of the equilibria}

There are two main constraints on the existence of an equilibrium for any set of loop parameters ($p_{0i}, B_{0i},T_i$) for a star of a given mass, radius and rotation rate. The first relates to the variation of the pressure along the loop. Clearly,  the magnetic pressure inside the loop must always be positive. From (\ref{press_bal}), however, we can see that the loop pressure is determined by the pressure of the external field and by the difference in the plasma pressure inside and outside the loop. While $B_e^2$ will always be positive, if ($p_e - p_i$) is sufficiently large and negative there may be no equilibrium for the loop.

We show in Fig. (\ref{pressure}) the variation of pressure with height for a loop whose temperature is equal to, lower than, or greater than its surroundings. In all cases, the pressure initially falls (at a rate that increases as the temperature falls) then starts to rise at the co-rotation radius where $g=0$ and hence from (\ref{gravity}) 
\begin{equation}
\frac{y_K}{R_\star}=\left( \frac{GM_\star}{R_\star^3 \omega^2}\right)^{1/3} -1. 
\end{equation}
The pressure continues to rise with height until it is equal to its base value at a height 
\begin{eqnarray}
\frac{y_m}{R_\star} & = & \frac{1}{2}\left(
                                             -3+\sqrt{1+\frac{8GM_\star}{R_\star^3 \omega^2}}
                                                       \right) \\
                               & = & \frac{1}{2}\left(
                                             -3+\sqrt{1+8 \left[
                                                                       \frac{y_K}{R_\star} + 1
                                                                  \right]^3}
                                                       \right).
\end{eqnarray} 
In Fig. (\ref{pressure}) we have chosen stellar parameters appropriate for AB Dor ($P_{\rm rot} = 0.514$ days) and so the co-rotation radius is at $1.7R_\star$ above the stellar surface. For clarity, the base pressures of the internal and external regions have been set equal, but changing them would only scale the curves up or down. For the case where the base plasma pressures are equal, the point $y_m$ is also where the pressure difference $(p_e - p_i)$ goes to zero and hence from (\ref{press_bal}) $B_i=B_e$. 
The effect of a stellar wind would be to reduce the external plasma pressure and hence to increase the size of the region in which the buoyancy force is negative.

We show also the pressure difference $(p_e - p_i)$ as a function of height for these cases. The pressure difference always has a turning point at the co-rotation radius, but for low internal temperatures, two other turning points are possible. One of these is close to the stellar surface, and one is close to $y_m$. Beyond $y_m$ the pressure difference also changes sign. 

The second constraint on the existence of an equilibrium comes from the balance of forces normal to the loop. As shown in (\ref{shape}) the loop must adjust its shape ($y'$) such that its Lorentz force is zero. The gradient of the magnetic pressure normal to the loop (the left-hand side of (\ref{shape})) must exactly balance the magnetic tension (the right-hand side of (\ref{shape})). The equilibrium shape of the loop therefore depends entirely on the nature of $B_i^2$ and hence, from (\ref{press_bal}) on the form of the external field $B_e$ and the pressure difference across the boundary of the loop $(p_e - p_i)$. The nature of this equilibrium is most easily understood by considering the region close to the loop summit, $y=H$ where the normal vector lies in the direction of $-{\bf y}$. Since the magnetic tension force acts inwards, the gradient of the magnetic pressure must act outwards, i.e. we must have $\partial B_i^2/\partial y <0$. This places significant restrictions on the types of loops that can reach an equilibrium, since pressure balance across the loop (\ref{bisq}) and hydrostatic equilibrium (\ref{dpbydy}) give
\begin{equation}
 \pderiv{B_i^2}{y} = \pderiv{B_e^2}{y} + 2 \mu (\rho_e - \rho_i) g.
 \label{condition2}
\end{equation}
Thus, at the loop summit, the downwards pull of tension is balanced by a combination of the pressure gradient of the external field and the buoyancy of the gas inside the loop. This buoyancy term may act outwards or inwards however as can be seen in the lowest panel of Fig. (\ref{pressure}). For loops that are just a little cooler than their environment, the buoyancy changes sign only at the co-rotation radius, but for loops that are significantly cooler than their neighbours, two additional changes of sign are possible. 

\section{Specific example}
In order to give an example of the types of loop equilibria that may be found, we choose a simple external field that is closed at low heights, then opens up beyond some given height $y_s$. For simplicity, we choose a field that is potential and two-dimensional, so we ignore any field components that lie in the north-south direction and consider a field structure that lies in the equatorial plane where the effective gravitational acceleration is purely radial. We choose the form
\begin{equation}
B_{ex} + i B_{ey}  = B_0 \sqrt{e^{2ikz} + e^{-2ky_s}}
\end{equation}
where $z = x+iy$ and $B_0^2=B^2(0,0) = 1+e^{-2ky_s}$ . As illustrated in the left-hand panel of Fig. (\ref{cartoon}) this field structure has a current sheet that extends from $y=y_s$ to infinity.  Given the two constraints on the existence of a loop equilibrium:
\begin{eqnarray}
B_i^2 = B_e^2 + 2 \mu (p_e-p_i) &  > & 0, \\
\left[ \pderiv{B_i^2}{y} \right]_{y=H} =\left[  \pderiv{B_e^2}{y} + 2 \mu (\rho_e - \rho_i) g \right]_{y=H} & < & 0
\end{eqnarray}
\begin{figure}
\begin{center}

\includegraphics[width=2.8in]{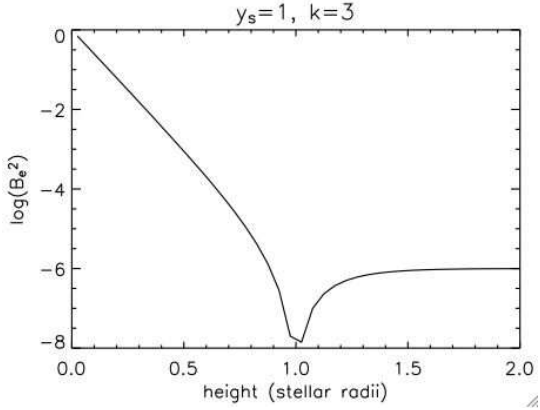}

 \caption{Variation with height of the external field strength squared (scaled to its base value). Values are calculated at the positions of the loop summits, i.e. at $x=\pi /2k$ and with the source surface at $y_s=1R_\star$ and  $k=3$. }

  \label{Bext}
 \end{center}
\end{figure}
it is helpful to consider how $B_e^2$ varies with height at the loop summits (i.e at $x=\pi /2k$). Here, $B_e^2 = e^{-2ky} -e^{-2ky_s}$ and so falls to zero when $y=y_s$. As shown in Fig. (\ref{Bext}), it rises beyond $y=y_s$ and tends to $B_e^2 = e^{-2ky_s}$ at large heights. Hence, below the source surface, the pressure gradient of the external field acts upwards and so acts to counteract the loop's own tension force. Thus  loops of both positive and negative buoyancy may be supported, provided the buoyancy force is not too large. Above the source surface, however, the pressure gradient of the external field acts downwards (in the same direction as the tension in the loop) although its magnitude tends to zero quite quickly. In this region, the only loops that can reach an equilibrium are those where buoyancy acts outwards and is sufficiently large to balance the tension force. Thus cool, dense loops are the most likely to find support above the source surface. As can be seen from Fig. (\ref{pressure}), however, the range of heights over which buoyancy is negative at the loop summit shrinks as the loop temperature drops. For very cool loops it is confined to a region close to the value of $y_m$. 

Clearly, for a given base pressure, the loop temperature is crucial in determining how buoyant the loop is and therefore the range of loop heights that give an equilibrium. Another crucial factor however is the stellar rotation rate since this determines the height of the co-rotation radius where $g=0$. If the source surface is outside the co-rotation radius, then loop equilibria may be found out for both negatively and positively buoyant loops out to the source surface. Above the source surface, loops where buoyancy acts outwards may also be supported. If, however, the source surface is inside the co-rotation radius, then there may be a range of heights between the source surface and the co-rotation radius where no solutions are possible because of the downward pressure of the external field at the loop summit. At the co-rotation radius itself, no solutions will be possible because the buoyancy force is zero here. Some solutions may be found at larger heights however, where the pressure of the external field has reduced and an outward buoyancy can support the loop against tension.

In order to clarify the types of loop equilibria that are possible, we consider a specific example. We choose the young solar-like star AB Dor, for which the co-rotation radius is at 1.7R$_\star$ above the stellar surface, and we choose an ambient coronal temperature T$_e =  2\times 10^7$K. Fig (\ref{single_shape}) shows the structure of some sample cool loops that extend above the source surface into the open field region. Clearly, any loops that form out in the open field region must have footpoint separations that are similar to that of the external arcade in which their lower sections are embedded. This immediately suggests that the range of loop geometries that is available for loops below the source surface is not available for those that form above it.
\begin{figure*}
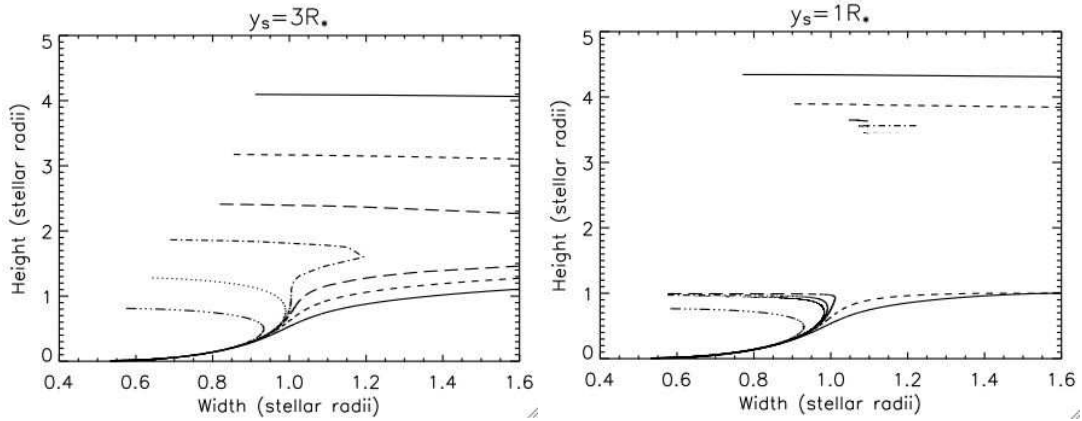

\begin{center}

\includegraphics[width=2.8in]{jardinefig7a.epsf}
\includegraphics[width=2.8in]{jardinefig7b.epsf}

 \caption{Loop height plotted against footpoint separation for loops at temperatures of: $2.5\times 10^7$K (dot-dot-dot-dash);  $9.1\times 10^6$K (dot); $8.7\times 10^6$K (dot-dash); $8.3\times 10^6$K (long dash); $7.1\times 10^6$K (short dash); $4.2\times 10^6$K (solid) . The source surface $y_s=3R_\star$ (left) and $y_s=1R_\star$ (right). In all cases, the plasma beta at the stellar surface is 0.01, the external coronal temperature is $T_e=2\times 10^7$K, the co-rotation radius is at $1.7$R$_\star$ and we choose $k=3$ so that the external arcade spans about 60$^\circ$ longitude at the stellar surface. }

  \label{bisq}
 \end{center}
\end{figure*}

We show in Fig. (\ref{bisq}) the heights and footpoint separations of loops at different temperatures for two cases, one on the left where the source surface is at a large height ($3R_\star$) and where we expect to recover results similar to those of \scite{jardine91loop} since the opening-up of the field should have a minimal effect on most of the volume of the corona. The other case, shown on the right of Fig. (\ref{bisq}) is where the source surface is placed at ($1R_\star$) and is therefore much closer to the stellar surface and affects much more of the coronal volume. 

If the source surface is at large heights, then most of the loops that may form have their summits below the source surface, within the closed field region. In this case, as shown in the left-hand panels of Fig. (\ref{bisq}), equilibrium solutions are available that are qualitatively similar to those found by \scite{jardine91loop}.  For a given loop temperature there are typically two types of solution: a low-lying case where the loop follows the path of the external field and a solution with a much greater summit height where the buoyancy of the gas within the loop significantly influences its shape. 

If the source surface is at much lower heights, and particularly if it is below the co-rotation radius, then, as shown in the right-hand panels of  Fig. (\ref{bisq}), the nature of the solutions is somewhat different. Below the source surface, solutions can be found that are similar to those in \scite{jardine91loop}. 
Above the source surface, only those loops whose summits have buoyancy acting outwards can be supported. Below the co-rotation radius, where $g<0$, this requires that $\rho_i<\rho_e$. At the co-rotation radius itself, $g=0$ and no solutions are possible. Above the co-rotation radius, $g>0$ and we need to have $\rho_i>\rho_e$. In addition, the magnitude of the buoyancy force must be large enough to balance the magnetic tension at the loop summit. This means that only loops with a large enough value of $\beta$ (the ratio of plasma to magnetic pressures) will be able to reach an equilibrium. If $\beta=0$ then no loop solutions will be possible above the source surface.

\begin{figure}
\begin{center}

\includegraphics[width=3in]{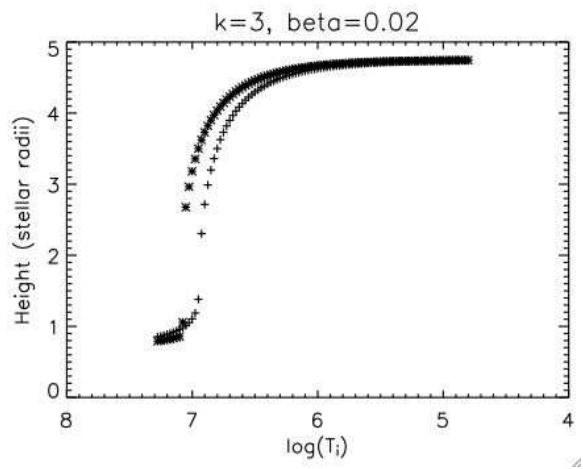}

 \caption{The maximum height of individual loops as a function of the temperature of the loop. The external corona is at a temperature of $T_e=2\times10^7$K and the star has a rotation period of 0.514 days, giving a co-rotation radius at $1.7$R$_\star$. The different symbols denote the height of the source surface $y_s$ which marks the extent of the closed corona: crosses denote loops where the source surface is above the co-rotation radius ($y_s=3$R$_\star$)  and stars denote loops where the source surface is below the co-rotation radius ($y_s=1$R$_\star$). }

  \label{maxheight}
 \end{center}
\end{figure}

The maximum height attainable by the loop is clearly a function of both temperature and the position of the source surface. As shown in Fig. (\ref{maxheight}), this maximum height increases with decreasing loop temperature. There is, however, a region around the co-rotation radius where there are no solutions.

\section{Discussion}

The aim of this paper is to demonstrate that for rapidly rotating stars, mechanical equilibrium exist for cool loops that have their summits well above the extent of the rest of the closed corona. The presence of co-rotating prominences trapped in the coronae of these stars therefore need not imply that their X-ray coronae extend out to the heights of 2-5R$_\star$ at which prominences are observed. This new model for the structure of stellar coronae is therefore capable of explaining both the recent X-ray observations that suggest that these coronae are compact, with the observations of prominences at heights well beyond that of the X-ray emitting gas.

The loop temperature is crucial in determining whether loops can be supported out in the open field region. Loops that extend into the open field need the outward pull of centrifugal forces acting on the gas inside the loop to support the loop against its own magnetic tension. The loop temperature is therefore important as it determines the magnitude and sign of the buoyancy force. It also determines the plasma pressure of the gas inside the loop which must not be so large that the loop has to expand enormously to remain in pressure balance with its surroundings. If the source surface is inside the co-rotation radius, then there may be a range of heights above the source surface where there are no solutions. In particular, no solutions can be found at the co-rotation radius itself since here the buoyancy force is zero and so cannot balance the tension of the loop. For those cool loops where the buoyancy acts outwards at the loop summit, solutions may be recovered again at larger heights, just below the critical height $y_m/R_\star=1/2(-3+\sqrt{1+8[y_K/R_\star +1 ]^3})$ where the plasma pressure in the loop rises to its base value. 

This critical height $y_m$ represents the maximum height for all cool loops. As the stellar rotation rate increases, both $y_K$ and $y_m$ decrease, and the maximum loop height moves closer to the surface. The behaviour of the source surface is however much more difficult to determine, as the energy density of both the coronal gas and the magnetic field depend on the stellar rotation rate, but in a manner that is not well understood. \cite{schrijver_asterosphere_03} find that the source surface should increase rapidly with rotation rate, up to activity levels perhaps 10 times that of the Sun. The behaviour of the source surface for the very rapid rotators discussed here is not clear however. It may well be that observing the distribution of prominence positions is one of the best ways of locating the source surface.

This simple model does not address the question of the stability of stellar prominences. To study the stability we would need to know more about their fine structure. Future observational studies should determine the filling factor of cool material in these prominences.
While the equilibria presented in this paper demonstrate the concept that cool loops can be supported within the open field regions beyond a star's close corona, they are not intended to model the structure of prominences themselves. The cool gas contained in the observed stellar prominences does not extend all the way to the stellar surface and the thin flux tubes of this model are unlikely to be a good approximation for stellar prominences, unless they are composed of many fine cool threads. A more sophisticated model would be required if the full thermal and spatial structure of prominences is to be understood. These cool equilibria may however act as the starting points for prominence formation. The range of observed prominence positions for AB Dor (between 2 and 5 stellar radii above the surface) is certainly consistent with the the values of $y_K=1.7$R$_\star$ and $y_m=4.7$R$_\star$.

If this model is correct and the prominences do indeed lie above the cusps of helmet streamers,  then, as on the Sun,  their positions should be correlated with the locations of the neutral polarity lines at the surface. The number of prominences would then be related to the number of changes of polarity at the surface, which is a measure of the complexity of the field. On the Sun, the strength of the small-scale field relative to the large-scale dipolar component varies through the solar cycle. It is smallest at cycle minimum where the field most closely resembles a dipole, and grows though the cycle as progressively more and more East-West bipoles emerge through the surface. It is therefore possible that if the prominence locations are related to surface polarity changes, then a systematic change in their number may signal changes in the underlying stellar field.

\section{Acknowledgements}

The authors acknowledge useful discussions with Andrew Collier Cameron and Nick Dunstone.





\end{document}